# A propensity-score integrated approach to Bayesian dynamic power prior borrowing

Jixian Wang[a] [*], Hongtao Zhang[b], Ram Tiwari[c]

[a]Bristol Myers Squibb, Boudry, Switzerland;

[b]Merck & Co., Inc., North Wales, Pennsylvania, USA

[c]Bristol Myers Squibb, Berkeley Heights, New Jersey, USA

October 5, 2022

**Abstract**

Use of historical control data to augment a small internal control arm in a randomized control trial (RCT) can lead to significant improvement of the efficiency of the trial. It introduces the risk of potential bias, since the historical control population is often rather different from the RCT. Power prior approaches have been introduced to discount the historical data to mitigate the impact of the population difference. However, even with a Bayesian dynamic borrowing which can discount the historical data based on the outcome similarity of the two populations, a considerable population difference may still lead to a moderate bias. Hence, a robust adjustment for the population difference using approaches such as the inverse probability weighting or matching, can make the borrowing more efficient and robust. In this paper, we propose a novel approach integrating propensity score for the covariate adjustment and Bayesian dynamic borrowing using power prior. The proposed approach uses Bayesian bootstrap in combination with the empirical Bayes method utilizing quasi-likelihood for determining the power prior. The performance of our approach is examined by a simulation study. We apply the approach to two Acute Myeloid Leukemia (AML) studies for illustration.

---
[*]CONTACT Jixian Wang: jixian.wang@bms.com

# 1 Introduction

In situations when the control arm of a randomized clinical trial (RCT) is smaller than the test arm in order to treat more patients with the test treatment, in order that statistical inference for treatment comparison is not compromised due to the small control arm, use of external data, which may be from another clinical trial or real-world data (RWD), has become a valuable source for augmenting the internal control of the RCT. This approach is often referred to as borrowing external controls. Regulatory guidance documents on using RWD, to aid drug development, have been published (EMA, 2020; FDA 2018). However, the use of external data also introduces the risk of potential bias, as the historical control population may be rather different from the RCT.

There are several approaches to eliminate or reduce the bias due to population difference. In particular, robust adjustment for the population difference using propensity score (PS) based approaches such as the inverse probability weighting or matching can make the borrowing more efficient and robust to model misspecification to some extent (Rosenbaum & Rubin, 1983; Robins et al., 1994). If we assume that the adjustment can eliminate the bias, one may be tempted to use a single arm trial with adjusted external control completely.

However, some assumptions such as no unobserved confounders can not be verified based on the data. Therefore, using adjusted external control has potential risk of introducing confounding bias.

An RCT, with a small control arm, may provide an internal reference for evaluating the difference between the internal and external control populations. To deal with the population difference, power prior approaches (Ibrahim et al., 2000, 2003; Hobbs et al., 2011, 2013; Neuenschwander et al. 2009) can be used to discount the historical data to mitigate the impact of the bias. The amount of borrowing can be either fixed or determined by discounting the historical data, based on the similarity of the outcomes of the two populations, known as Bayesian dynamic borrowing. However, a considerable difference may still exist and lead to a moderate bias. To mitigate the impact of population difference, some recently developed approaches used PS matching or stratification to reduce the difference within matched pairs or strata, and then applied the power prior within them (Wang et al., 2019a, 2019b; Sachdeva et al., 2021). An alternative approach used a linear outcome model assuming exchangeability after covariate adjustment (Kotalik et al., 2021).

As a further development based on the above-mentioned work, we propose a novel approach integrating PS based approaches for the covariate adjustment and Bayesian dynamic borrowing using the power prior (Ibrahim et al., 2000, 2003; Hobbs et al., 2011, 2013; Chen et al., 2011; Gravestock et al., 2017, 2018; Wang et al., 2019a, 2019b). The proposed approach combines the advantages of propensity score based approaches for adjusting confounding bias without specifying the outcome model, and the power prior that down-weights the information from the historical data if, after adjustment, it is still considerably different from the internal control. Our approach is an approximate full Bayesian that takes the uncertainty of model fitting and weighting into account.

The major challenge is that the PS based methods are frequentist approaches with minimum model specification, while the power prior methods are built in the Bayesian framework. Our approach is partially built on the work of approximate PS-based Bayesian approaches (Graham et al., 2016; Saarela et al., 2016; Capistrano et al., 2019), in which the inverse probability weighting (IPW) and doubly robust (DR) estimation approaches were put in the Bayesian framework and the posterior distribution was approximated by Bayesian bootstrap (BB) (Rubin, 1981). Our approach is considerably simpler than the outcome based full Bayesian approach using MCMC such as Kotalik et al. (2021).

## 2 A review of relevant approaches

### 2.1 Borrowing historical controls to augment an internal control arm

First, we state our approach of borrowing historical controls to augment an internal control arm formally. Let $D_i = (y_i, \mathbf{X}_i, H_i), i = 1, ..., n$, be the outcome, covariates, and an indicator of being in the historical control for the $i$th subject in the combined of the internal trial population and the historical control population. The sample sizes and the means of the internal and historical control populations are: $n_h = \sum_{i=1}^{n} H_i, n_0 = n - n_h$, $\bar{y}_0 = \sum_{i=1}^{n}(1-H_i)y_i/n_0$ and $\bar{y}_h = \sum_{i=1}^{n} H_i y_i/n_h$, respectively. We also denote the whole dataset as $D = (D_1, ..., D_n)$, and those of the internal and external controls as $D^0 = D|H_i = 0$ and $D^h = D|H_i = 1$, respectively. Although our final goal is to evaluate the treatment effect of the treatment applied in the treated arm in the trial population, the key issue we concentrate on is the evaluation of treatment effect under the control: $\mu = E(y_i|H_i = 0)$, borrowing historical control data with confounding adjustment. Here, $y_i$ could be either a continuous or binary variable, in the

latter case, $\mu$ is the rate or proportion of the outcome. Due to population difference between the trial and historical control, $E(\bar{y}_h) = E(y_i|H_i = 1)$ is likely different from $\mu$, hence adjustment for population difference is often necessary.

## 2.2 Propensity score based adjustment

A commonly used approach for population adjustment is based on PS defined as the probability of belonging to the historical control, given the covariates $\mathbf{X}_i$:

$$e_i = P(H_i = 1|\mathbf{X}_i) \qquad (1)$$

which is often modelled by a logistic regression as $P(H_i|\mathbf{X}_i, \gamma)$ with parameters $\gamma$. Under some technical conditions, $y_i \perp H_i|e_i$, hence we can use the inverse probability waiting (IPW) estimator

$$\hat{\mu}_{ipw} = (\sum_{i=1}^{n} H_i w_i)^{-1} \sum_{i=1}^{n} H_i w_i y_i \qquad (2)$$

where $w_i = (1 - e_i)/e_i$, to estimate $\mu$. Using the property $y_i \perp H_i|e_i$, it is straightforward to show that $E(\hat{\mu}_{ipw}) = \mu$. Therefore, when the PS model is correctly specified, one can combine $\hat{\mu}_{ipw}$ with the internal control mean $\bar{y}_0$ for more accurate estimation of $\mu$. This IPW estimator (2) is slightly different from the standard IPW estimator for the average treatment effect (ATE) for the whole population, since we aim at estimating the control effect in the trial population.

## 2.3 Propensity score based Bayesian methods

Although PS based approaches were proposed from frequentist aspect, effort has been made to use PS in the Bayesian framework to provide a robust Bayesian approach for population adjustment (Zigler, 2016; Zigler et al., 2013, 2014).

McCandless et al. (2009) used joint models for the PS and outcome, and later included the PS as a covariate (McCandless et al., 2010).

An alternative to the modeling approach uses the IPW estimator and resampling to obtain the posterior distribution of $\mu$ (Graham et al., 2016, Capistrano et al., 2019). Capistrano et al. proposed an approach using BB (Rubin, 1981; Newton and Raftary, 1994):

1. Repeat steps 2-4 (below), for $s = 1, ..., S$, times.

2. Generate $\xi_{si}, i = 1, ..., n$, independently, from the uniform Dirichlet distribution.

3. Fit the PS model (1) for $H_i$ using weights $\xi_{si}$.

4. Obtain
$$\hat{\mu}_{ipws} = (\sum_{i=1}^{n} H_i w_i \xi_{is})^{-1} \sum_{i=1}^{n} H_i \xi_{is} w_i y_i \tag{3}$$

The $S$ estimates $\hat{\mu}_{ipws}$ can be considered as posterior samples of $\mu$. An in-depth consideration of their properties, including formulating this approach in the Bayesian framework using de Finetti's representation (de Finetti, 1974) can be found in Saarela et al. (2016).

## 2.4 Power prior and Bayesian dynamic borrowing

Although the previous approaches can provide adjusted estimates of $\mu$ in a frequentist or Bayesian framework, combining them with $\bar{y}_0$ remains a challenge, especially when the IPW adjustment may not completely eliminate the confounding bias due to population difference. For mitigating the potential bias, Bayesian borrowing with the power prior is a powerful tool for this purpose (Ibrahim et al., 2000, 2003; Hobbs et al., 2011, 2013; Chen et al., 2011; Gravestock et al., 2017, 2018; Wang et al., 2019a, 2019b). Let $D^0$ and $D^h$ denote data from the internal and historical control data, respectively, $\theta$ denote model parameters, and $L(\theta|D)$ denote the likelihood function given data $D$. Our goal is to estimate $\mu$, the mean response given control data $D^0$ and $D^h$. The power prior, conditional on $D^h$, is formulated as

$$\pi(\theta|D^h, a_0) \propto L(\theta|D^h)^{a_0} \pi_0(\theta) \qquad (4)$$

where $0 \leq a_0 \leq 1$ is the power prior (discounting) parameter in the likelihood of historical data, and $\pi_0(\theta)$ is the initial prior for $\theta$. The corresponding posterior distribution is

$$\pi(\theta|D^h, D^0, a_0) \propto L(\theta|D^0) L(\theta|D^h)^{a_0} \pi_0(\theta).$$

The parameter $a_0$ allows one to control the weight of historical data in the current study. One can choose $a_0$ close to zero when historical and current data are highly heterogeneous, and choose $a_0$ close to one when they are congruent. Here $a_0$ is taken as fixed and a sensitivity analysis can be carried out to determine an appropriate value of $a_0$.

Assume that $\bar{y}_h$ and $\bar{y}_0$ are normally distributed with common mean $\mu$, and variances $\sigma_h^2$ and $\sigma_0^2$ in the historical and internal control populations, respectively. With fixed $a_0$, the posterior distribution of $\mu$ is

$$\mu \sim N(\hat{\mu}, \hat{\sigma}^2) \qquad (5)$$

where

$$\hat{\mu} = \hat{\sigma}^2 (a_0 \sigma_h^{-2} \bar{y}_h + \sigma_0^{-2} \bar{y}_0) \qquad (6)$$

$$\hat{\sigma}^2 = (a_0 \sigma_h^{-2} + \sigma_0^{-2})^{-1} \qquad (7)$$

In applications, $\sigma_h^2$ and $\sigma_0^2$ can be replaced by the corresponding sample variances.

One can also consider $a_0$ as random, with a non-informative prior. Then, its distribution, consequently the amount of borrowing, will depend on the similarity between $D^h$ and $D^0$. A simple approach to estimate it utilizes EB approach (Gravestock and Held, 2017, 2018). This approach estimates $a_0$ by maximizing the marginal likelihood $L(a_0|D^0, D^h)$. For the normal distribution case above, it has a closed form:

$$\hat{a}_0 = \frac{\sigma_h^2}{\max[(\bar{y}_h - \bar{y}_0)^2, \sigma_h^2 + \sigma_0^2] - \sigma_0^2} \tag{8}$$

For binomial outcomes, with a $Beta(1,1)$ prior and a fixed $a_0$, the posterior distribution of $\mu$ can be written as

$$\mu \sim Beta(a_0 y_{h.} + y_{0.} + 1, n_0 + a_0(n_h - y_{h.}) - y_{0.} + 1) \tag{9}$$

where $y_{h.} = \sum_{i=1}^n H_i y_i$ and $y_{0.} = \sum_{i=1}^n (1 - H_i) y_i$.

Assuming a random $a_0$, its marginal likelihood is (Eq 4, Gravestock and Held, 2018)

$$\pi(a_0|D^h, D^0) \propto \frac{Beta(a_0 y_{h.} + y_{0.} + 1, a_0(n_h - y_{h.}) + n_0 - y_{0.} + 1)}{Beta(a_0 y_{h.} + 1, a_0(n_h - y_{h.}) + 1)} \tag{10}$$

To determine $a_0$, we can use the approach of Gravestock and Held (2017) and find

$$a_0^* = \mathrm{argmax}_{a_0} \pi(a_0|D^h, D^0) \tag{11}$$

within the range of [0,1]. In practice, this can be done with a grid search.

## 2.5 Bayesian borrowing with covariate adjustment

The extent of borrowing in the dynamic approach (Gravestock et al., 2017) depends on the difference in observed mean responses between internal and external control groups. A part of the difference may be due to the difference in $\mathbf{X}_i$ between internal and external controls. In order to reduce such difference, some approaches based on the PS have been proposed. Wang et al. (2019a, 2019b) proposed a stratification based on the PS and using the overlapping area between the PS distributions of subjects of historical and internal controls. This approach uses all subjects from the historical control who meet inclusion/exclusion criteria and discounts those with a large difference to mitigate their impact. The amount of borrowing from individuals is determined by the similarity in the PS distributions, rather than the outcomes, between the internal and external controls.

# 3 A PS integrated approximate Bayesian approach

Our integrated approach combines posterior sampling using BB, PS adjustment with the IPW estimator, and the EB estimator for the power prior parameter $a_0$. One advantage is that $a_0$ is not fixed, and a posterior distribution can be obtained with BB. We use BB to generate posterior samples of the mean and variance for both internal and historical controls. For the latter, we adapt the IPW estimator (2) as did in Capistrano et al. (2019) to adjust for confounding bias, and the justification of combining the BB weights with the IPW weights as described in Section 6 of Saarela et al. (2016). We start with introducing some general notation here. Let $l(\boldsymbol{\theta}, D_i)$ be the log-likelihood function for $D_i$ with parameters $\boldsymbol{\theta}$. Following Saarela et al. (2016), we maximize $E(l(\boldsymbol{\theta}|D_i)|\mathbf{D})$,

which can be estimated by BB as:

$$E(l(\boldsymbol{\theta}|D_i)|\mathbf{V}) \approx \sum_{i=1}^{n} w_i(\xi)\xi_i l(\boldsymbol{\theta}|V_i) \tag{12}$$

where $\xi_i \sim Dirichlet(1,...,1)$ are the BB weights, $\xi = (\xi_1, ..., \xi_n)$ is a full set of the weights, and $w_i(\xi)$ is the IPW weight, depending on the BB weights as well. Then $\boldsymbol{\theta}$ is estimated as

$$\hat{\boldsymbol{\theta}} = \text{argmax}_\theta (\sum_{i=1}^{n} w_i(\xi)\xi_i l(\boldsymbol{\theta}|D_i)) \tag{13}$$

In our case, $l(\boldsymbol{\theta}|D_i)$ can be decomposed into two parts $l(\boldsymbol{\theta}|D_i) = p(y_i|\mu)p(H_i = 1|\mathbf{X}_i, \boldsymbol{\gamma})$ with $\boldsymbol{\theta} = (\mu, \boldsymbol{\gamma})$. Therefore, the two sets of parameters can be estimated separately, given the BB weights $\xi$. For a detailed discussion on the impact between the outcome and propensity score models, see Saarela et al. (2016).

To combine this BB approach with the EB method introduced in Section 2.4, we take a two-step approach. For each set of BB weights $\xi$, the first stage uses the BB approach to obtain the weighted (and adjusted) means $\hat{y}_h$ and $\hat{y}_0$, for both external and internal controls, respectively. For the latter, we weight $y_i$s of internal control by BB weights

$$\hat{y}_0 = (\sum_{i=1}^{n}(1-H_i)\xi_i)^{-1} \sum_{i=1}^{n}(1-H_i)\xi_i y_i. \tag{14}$$

The BB weighting is not necessary for the internal controls, even adjustment is not needed, to obtain the approximate posterior samples for $\mu$. $\hat{y}_h$ is an adjusted mean of external controls, which can, but not necessarily, be estimated as in (3). Although both $\hat{y}_h$ and $\hat{y}_0$ depend on $\xi$, we will suppress $\xi$ in the notation for simplicity.

The second step is based on a quasi log-likelihood conditional on a BB real-

ization $\xi$:

$$l(\mu, a_0|\xi) \propto [(\hat{y}_0 - \mu)^2/\hat{\sigma}_0^2 + a_0(\hat{y}_h - \mu)^2/\hat{\sigma}_h^2]/2 \qquad (15)$$

where $\hat{\sigma}_0^2$ and $\hat{\sigma}_h^2$ are the weighted sample variances of $\hat{y}_0$ and $\hat{y}_h$, respectively, and we denote it as $l(\mu, a_0|\xi)$ to emphasize its dependence on the BB weights via $\hat{\sigma}_0^2, \hat{\sigma}_h^2$, $\hat{y}_0$ and $\hat{y}_h$. For example, $\hat{\sigma}_0^2 = \sum_{i=1}^{n}(1 - H_i)\xi_i^*(y_i - \hat{y}_0)^2/(n_0 - 1)$ and $\xi_i^*$ is normalized $\xi_i$ such that $\sum_{i=1}^{n}(1 - H_i)\xi_i^* = n_0$. $\hat{\sigma}_h^2$ is calculated in the same way as for $\hat{\sigma}_0^2$ using normalized $\xi_i w_i(\xi)$. In our simulation and the analysis below, we use R-function wtd.var(.) in the Hmisc package (Harrell, 2021). Eq (15) has the same form as Eq 4 of Gravestock and Held (2018), which leads to (8) and (6). Therefore, replacing $\bar{y}_h, \bar{y}_0$ with $\hat{y}_h, \hat{y}_0$, $\sigma_0^2, \sigma_h^2$ with $\hat{\sigma}_0^2$ and $\hat{\sigma}_h^2$ in (8), we have

$$\hat{a}_0 = \frac{\hat{\sigma}_h^2}{\max[(\hat{y}_h - \hat{y}_0)^2, \hat{\sigma}_h^2 + \hat{\sigma}_0^2] - \hat{\sigma}_0^2}. \qquad (16)$$

Replacing $a_0$ with $\hat{a}_0$ in (6) and (7), we obtain a BB sample of $\mu$.

Algorithm 1 gives steps of the integrated algorithm for normally distributed outcome. It repeats the above steps $s = 1, ..., S$ times to obtain a posterior sample of $\mu$.

---

**Algorithm 1** Bayesian bootstrap for posterior sample of $\mu$ with normal outcomes.

---
**Require:** $Y_i, X_i, H_i, S$
1: **for** Bootstrap run $s = 1$ to $S$ **do**
2:     Generate weights $\xi_{si}, i = 1, ..., n$ from the uniform Dirichlet distribution $Dirichlet(1, ..., 1)$
3:     Fit a PS model for $H_i$ including $\mathbf{X}_i$ and using weights $\xi_{si}$.
4:     Obtain $\hat{y}_{hs} = \hat{\mu}_{ipws}$ weighted by $\xi_{si}$s as in (3), and its weighted sample variance $\hat{\sigma}_{hs}^2$.
5:     Obtain weighted mean $\hat{y}_{0s}$ and sample variance $\hat{\sigma}_{0s}^2$ using weight $\xi_{is}$.
6:     Replace $\bar{y}_0, \bar{y}_h$ and $\sigma_0^2, \sigma_h^2$ with $\hat{y}_{0s}, \hat{y}_{hs}$ and $\hat{\sigma}_{0s}^2, \hat{\sigma}_{0s}^2$ in (8) to obtain $\hat{a}_{0s}$
7:     Replace $\bar{y}_0, \bar{y}_h$ and $\sigma_0^2, \sigma_h^2$ with $\hat{y}_{0s}, \hat{y}_{hs}$ and $\hat{\sigma}_{0s}^2, \hat{\sigma}_{0s}^2$, and $a_0$ with $\hat{a}_{0s}$ in (7) and (6) to obtain $\hat{\mu}_s$ from (6).
8: **end for**
9: Output $\hat{\mu}_1, ..., \hat{\mu}_S$.

---

Then, $\hat{\mu}_s, s = 1, ..., S$, can be considered as approximate posterior samples of $\mu$. Steps 2-4 of the algorithm are similar to those in Capistrano et al. (2019). Steps 5-7 are for Bayesian borrowing with $a_0$ determined by empirical Bayesian, incorporated into the full BB steps. Note that $\hat{\mu}_s$ is weighted by both the PS and the BB weights, although the PS is already a bootstrapped version. This is necessary for using BB for approximate Bayesian inference. See Saarela et al. (2016) for technical details.

The above algorithm can be easily adapted for binomial outcomes. The major difference is in steps 5-7, which should be replaced by:

5: In (10) replace $y_{h\cdot}$ with $n_h \hat{y}_{hs}$ and $y_{0\cdot}$ with $n_0 \hat{y}_{0s}$.

6: Obtain $\hat{a}_{0s}$ from (11) with a grid search. In this paper, we search using a 0.02 grid over the range [0,1].

7: Obtain
$$\hat{\mu}_s = \frac{\hat{a}_{0s} n_h \hat{y}_{hs} + n_0 \hat{y}_{0s} + 1}{\hat{a}_{0s} n_h + n_0 + 2}. \quad (17)$$

In the last step, the posterior mean is taken as $\hat{\mu}_s$.

The approach enjoys some properties from both IPW adjustment and Bayesian dynamic borrowing with power priors and BB, which ensures some asymptotic properties. Here, we give a short summary, with intuition based on (16) without technical details. With a large sample size $(\hat{y}_h - \hat{y}_0)^2 = \sigma_h^2 + \sigma_0^2 + \delta^2$, where $\delta$ is the true difference between the internal and adjusted external means. When there is no adjustment or the adjustment is invalid, $\delta^2$ does not tend to zero, but the two variances do. This leads to $\hat{a}_0$ close to zero, meaning no borrowing to avoid biases. In contrast, with a valid adjustment, $\delta^2$ will be close to zero, hence (16) is approximately $\sigma_h^2 / [(\sigma_h^2 + \sigma_0^2) - \sigma_0^2] = 1$, leading to almost full borrowing. Therefore, in the proposed approach, the use of IPW or another adjustment, can improve fixed or dynamic borrowing, and the dynamic borrowing also provides safeguard against invalid frequentist confounding adjustment.

## 4 A simulation study

A simulation study is conducted to examine the performance of the integrated approach. The difference between the two control populations is represented by the mean difference in $\mathbf{X}_i$. The simulation data generating models and parameters are:

- $\mathbf{X}_i \sim N(0, \mathbf{I}_p)$ when $H_i = 0$ and $\mathbf{X}_i \sim N(\mathbf{b}, \mathbf{I}_p)$, where $\mathbf{I}_p$ is a p-dim identity matrix, and varying $\mathbf{b}$ represents the population difference between internal and historical controls. This leads to a logistic PS model.

- The outcome model is $Y_i = \boldsymbol{\beta}^T \mathbf{X}_i + \varepsilon_i$, where $\varepsilon \sim N(0, 1)$ hence the population difference leads to a difference $\boldsymbol{\beta}^T \mathbf{b}$.

- $\boldsymbol{\beta} = \beta \mathbf{1}_p$ and $\mathbf{b} = b \mathbf{1}_p$, where $\mathbf{1}_p$ is a $p$-vector of ones. We set $p = 5$ or $p = 10$, $\beta = 0.3$ and vary $b$. The last is the key parameter, as a measure of population difference.

- We run 1000 simulations. For each, 100 bootstrap runs are used to estimate the posterior distribution of $\mu$.

The simulation examines four estimators: Average differences with no and full borrowing, and dynamic borrowing with and without IPW adjustment. Without IPW adjustment, the above settings would lead to a negative bias of different sizes in the combined estimation of $\mu$.

Figure 1 shows the posterior distributions of $\mu$ of dynamic borrowing with and without IPW adjustment by mean difference of covariates between internal and historical controls $E(\mathbf{X}_i | H_i = 0) = 0$ and $E(\mathbf{X}_i | H_i = 1) = -b$, compared with those of no borrowing and full borrowing, with normally distributed outcomes for $p = 5$. When $b = 0$, the IPW adjusted estimator has slightly larger variability than the unadjusted one, due to the variability in $w_i$s. With increasing $b = 0.15$ and $b = 0.3$, the unadjusted estimate downward bias and increasing variability. The IPW adjusted estimates seemed unbiased with relatively lower variability than the unadjusted with all $b$ values, showing the benefit of IPW adjustment.

Table 1 gives mean bias, variance, MSE and variance ratio (relative to no borrowing) of four estimators with no and full borrowing, and dynamic borrowing with and without IPW adjustment for normal outcomes with $p = 5$ and $p = 10$. As expected, when there is no difference between the internal and external control populations, using IPW adjustment increases variability. However, when there is a difference, using IPW reduces not only the bias, but also the variability for $p = 5$. The results with $p = 10$ have a similar pattern, except that all estimators perform worse than those with $p = 10$. Particularly, when $b = 0.6$, the variance ratio is almost 1. In general, the bias of dynamic borrowing without IPW is larger than that with IPW, but still much smaller than the full borrowing bias (and the one with fixed borrowing with $a_0 = 0.5$, which is 1/2 of the full borrowing bias). This result is consistent with the properties we claim at the end of the section above, because dynamic borrowing without IPW can be considered as a surrogate scenario of no proper population adjustment.

A similar simulation has also been conducted for binary outcomes. The simulation setting is similar to those above, except that $Y_i \sim Bin(p_i)$ with $p_i = 1/(1 + \exp(-\boldsymbol{\beta}^T \mathbf{X}_i))$. The results with $p = 5$ and different $b$ values are presented in Figure 2, showing a similar pattern to that in Figure 1. . Table 2 gives mean bias, variance, MSE and variance ratio(vs no borrowing) of four estimators with no and full borrowing, and dynamic borrowing with and without IPW adjustment for binary outcome with $p = 5$ and $p = 10$. Similar patterns as those of normally distributed outcomes are found.

## 5 An illustrative example

We illustrate our approach using two publicly available data sets evaluating gemtuzumab ozogamicin (GO, a CD-33 targeted therapy) with chemotherapy for children and adolescents with acute myeloid leukemia (AML): AML03P1 (Cooper et al., 2012) and AML0531 (Gamis et al., 2014). Both trials had a GO with chemotherapy arm which consisted of a remission induction phase (course 1) followed by an intensification phase (course 2). We use patients who had the status of complete remission (CR) ascertained at the end of course 2. The number of patients are 59 and 234 respectively from trial AML03P1 and AML0531.

Our aim is to borrow data from AML0531 to strengthen the small arm in AML03P1. We use IPW to control the difference in the following baseline factors: log-age, log-bone marrow leukemic blast percentage (log-BM), central nervous system (CNS) disease, race, risk group and white blood count (WBC) count at diagnosis. Table 3 presents the fitted PS model together with the mean / percentage differences of the above covariates. There is a substantial difference in log-WBC and also smaller differences in race and high-risk. The weighted differences are generally much smaller, in particular in log-WBC, showing the effect of covariate balancing with IPW. To apply the proposed approach, the algorithm presented in Section 3 is implemented in R with the code given in the Appendix. Figure 3 shows the posterior density, median and 95% credible interval of the CR rate at the end of course 2 for AML03P1 with Bayesian dynamic borrowing with and without IPW adjustment, compared to those with full and no borrowing. The density with full borrowing is rather different from the original AML03P1 one (no borrowing), while the dynamic borrowing one is in between the two. The IPW adjusted one is more similar to the original one, but with less variability. The median(SE) of the four estimator(Bayesian dynamic borrowing without and with IPW, full and no borrowing) are 0.94(0.024), 0.95(0.017), 0.91(0.017) and 0.97(0.025), respectively. These results show the advantage of IPW adjustment before applying Bayesian dynamic borrowing when the difference between the trials is significant.

## 6 Discussion

We have proposed a novel approach integrating propensity score for the covariate adjustment and Bayesian dynamic borrowing using power prior. The approach combines the advantages of propensity score based approach for adjusting confounding bias without specifying the outcome model, and the power prior that down-weights the historical data if, after adjustment, it is still considerably different from the internal control. Our approach is an approximate full Bayesian that takes the uncertainty of model fitting and weighting into account. Our approach utilizes Bayesian bootstrap, in combination with the empirical Bayesian method for determining the power prior, and is easier to implement than a full Bayesian approach using MCMC. The simulation results showed robust performance of our approach under different scenarios and is generally a better approach than dynamic borrowing without adjustment.

One advantage of our approach for drug development is that the IPW step does not depend on the outcome data. Therefore, both IPW and BB weights can be determined without access to the outcome data and can be locked before the outcome database unlock. For IPW weights, one may change the PS model, e.g., when a convergence problem occurs, or truncate extreme weights if approperite, without accessing the outcome. For the BB weights, it would be sufficient to lock the seed of random number.

We have focused on using IPW adjustment in our approach here for simplicity. Nevertheless, our approaches can also use the DR adjustment. The DR estimator combines $\hat{\mu}_{ipw}$ with a prediction of $\mu$ using covariates:

$$\hat{\mu}_{dr} = (\sum_{i=1}^{n} H_i(1-e_i)e_i^{-1})^{-1} \sum_{i=1}^{n} e_i^{-1}[(1-e_i)H_iY_i - (H_i - e_i)m_0(\mathbf{X}_i, \hat{\boldsymbol{\beta}})] \quad (18)$$

where $m_0(\mathbf{X}_i, \hat{\boldsymbol{\beta}})$ is an outcome model for $y_i$ with estimated parameters $\hat{\boldsymbol{\beta}}$ such that

$$E(\sum_{i=1}^{n}(1-H_i)m_0(\mathbf{X}_i, \boldsymbol{\beta}))/n_0 = \mu$$

This estimator is DR, as it is consistent if either the PS model (1) or $m_0(\mathbf{X}_i, \hat{\boldsymbol{\beta}})$ is correctly specified. These frequentist approaches do not need specification of a full model, and hence are more robust. The use of BB approach based on $\hat{\mu}_{dr}$ has been proposed by Graham et al. (2016). It is clear that the DR estimator (18) uses two components: IPW and outcome prediction. BB approach based on outcome prediction has been well studied; therefore, we have omitted the DR approach. To use our approach, we apply the BB weights to fitting $m_0(\mathbf{X}_i, \boldsymbol{\beta})$ as well as (18).

Our approach aims at borrowing historical controls to augment internal control, although the final goal is to compare the active treatment arm. An alternative is to adjust all trial subjects to the overall trial population with, e.g., an analysis of covariance, then use the adjusted controls in the algorithms in Section 3. In particular, the IPW or DR approach to adjust historical controls will target the entire trial population, e.g., the PS model will be fitted with all trial subjects labeled as $H_i = 0$. Also note for approximate Bayesian interpretation, the BB approach should also include the analysis of covariance step.

Our approach is flexible enough to use other adjustment approaches such as other (not PS based) covariate balancing weights; for example, the calibration estimation, including the so called match adjusted indirect comparison in health economics. These approaches weight historical control patients to balance the covariates that are potential prognostic factors. Then use the same weights to obtain a weighted mean of the historical controls. As discussed above, our approach takes $\hat{\mu}_{ipw}$ as an adjusted estimator for $\mu$, then determines $a_0$ according to its similarity to the mean of internal controls. $\hat{\mu}_{ipw}$ can be replaced by $\hat{\mu}_{dr}$ and can also be replaced by the weighted mean of calibration estimation. One can also use an estimate with direct adjustment with an outcome model, but this approach will require outcome data in the internal control arm. Our approach can also be adapted to borrow from multiple sources of external controls. For example, for binary outcome, the approach can be combined with the method of Gravestock and Held (2018).

Direct adjustment using outcome models is another approach we have not mentioned, but can also be used together with the algorithm we proposed. With a fitted outcome model $m_0(\mathbf{X}_i, \hat{\boldsymbol{\beta}})$ to the internal control data with BB weights, one can use the predicted mean $\hat{y}_h = \sum_{i=1} H_i \xi^* m_0(\mathbf{X}_i, \hat{\boldsymbol{\beta}})/n_h$ in Section 3. This approach can be considered as a special case of the DR approach described above. We mention it separately, as it can also be used with a standard Bayesian method using a (generalized) linear model (Ibrahim et al., 2010, 2015). While the standard Bayesian approach may be more accurate, our approach relies on less assumptions, is more flexible and easy to use.

# 7 Appendix: R-code

```
    library(Hmisc)
simu_BBPS=function(nh=100,nsimu=1000, nboot=100,
                nc=100,np=5,b1=0.3,
                b=0.3){
  Bout=NULL
  EBout=NULL
  beta=rep(b1,np)
  for (simu in 1:nsimu){
    Xh=matrix(nrow=nh,rnorm(nh*np))-b
    Xc=matrix(nrow=nc,rnorm(nc*np))
    yh=Xh%*%beta+rnorm(nh)
    yc=Xc%*%beta+rnorm(nc)
    yv=c(yh,yc)
    hv=c(rep(1,nh),rep(0,nc))
    Xv=rbind(Xh,Xc)
    #EB+BB
    for (BB in 1:nboot){
      wi=rexp(nc)
      vi=rexp(nh)
      wi=wi/mean(wi)
      vi=vi/mean(vi)
      muc=mean(yc*wi)
      muh=mean(yh*vi)
      muf=sum(yc*wi+yh*vi)/sum(wi+vi)
      sigc=wtd.var(yc,wi)/nc
      sigh=wtd.var(yh,vi)/nh
      a0=sigh/(max((muc-muh)^2,sigc+sigh)-sigc)
      sig0=1/(1/sigc+ a0/sigh)
      mu0=(muc/sigc+a0*muh/sigh)*sig0
      fitps=glm(hv~Xv,family = "binomial",weights=c(vi,wi))
      ps=predict(fitps,type="response")[hv==1]
      odd=(1-ps)/ps*vi
      odd=odd/mean(odd)
      muc=mean(yc*wi)
      muh=mean(yh*odd)
      sigc=wtd.var(yc,wi)/nc
      sigh=wtd.var(yh,odd)/nh
      aps=sigh/(max((muc-muh)^2,sigc+sigh)-sigc)
      sig0=1/(1/sigc+ a0/sigh)
      mups=(muc/sigc+a0*muh/sigh)*sig0
      Bout=rbind(Bout,c(b,a0,aps,muc,muf,mu0,mups))
    }
  }
  #  list(Bout,EBout)
```

```r
    Bout
}

Qout=NULL
Qout=rbind(Qout,simu_BBPS(b=0))
Qout=rbind(Qout,simu_BBPS(b=0.15))
Qout=rbind(Qout,simu_BBPS(b=0.3))
Qout=rbind(Qout,simu_BBPS(b=0.6))

library(ggplot2)
ss=dim(Qout)[1]/4
bv=unique(Qout[,1])
Estimates=as.numeric(Qout[,4:7])
Estimator=rep(c("No borrowing","Full borrowing","Dynamic","IPW+Dynamic"),rep(ss*4,4))
Estimator=factor(Estimator,levels=c("Dynamic","IPW+Dynamic","No borrowing","Full borrowing")
vB=paste("b=",rep(Qout[,1],4))
Pout=data.frame(Estimates,b=vB,Estimator)
ggplot(data=Pout,aes(x=Estimator, y=Estimates)) + geom_boxplot()+facet_wrap(~b)

######################### AML example #################
aml2 <- aml1 %>% select(
  `CR status at end of course 2`,
  `MRD % at end of course 2`,
  `Age at Diagnosis in Days`,
  `Bone marrow leukemic blast percentage (%)`,
  `CNS disease`,
  `Protocol`,
  `Race`,
  `Risk group`,
  `WBC at Diagnosis`)
aml2$`MRD % at end of course 2` <- as.numeric(aml2$`MRD % at end of course 2`)
aml2 <- aml2[complete.cases(aml2),]
aml2$log_MRD <- log(aml2$`MRD % at end of course 2`+0.1)
aml2$log_age <- log(aml2$`Age at Diagnosis in Days`)
aml2$log_WBC <- log(aml2$`WBC at Diagnosis`)
aml2$log_BM <- log(aml2$`Bone marrow leukemic blast percentage (%)`+0.1)

# create binary variables
aml2$`CNS disease` <- ifelse(aml2$`CNS disease`=="No", 0, 1)
aml2$`Race` <- ifelse(aml2$`Race`=="Black or African American", 1, 0)
aml2$low_risk <- ifelse(aml2$`Risk group`=="Low", 1, 0)
aml2$high_risk <- ifelse(aml2$`Risk group`=="High", 1, 0)
aml2$curr=ifelse(aml2$Protocol == "AAML0531",1,0)
aml2$cr1=1*(aml2$`CR status at end of course 2` == "CR")
aml2$age=aml2$`Age at Diagnosis in Days`/365
```

```r
modelmat=as.matrix(aml2[,-c(1:3,6,8,16,17)])
nn=table(aml2$curr)
nc=nn[1]
nh=nn[2]
nbb=1000
Out=NULL
for(BB in 1:nbb){
  wi=rexp(nc)
  vi=rexp(nh)
  wi=wi/mean(wi)
  vi=vi/mean(vi)
  yc=aml2$cr1[aml2$curr==0]
  yh=aml2$cr1[aml2$curr==1]
  curr=c(rep(0,nc),rep(1,nh))
  modelmat2=rbind(modelmat[aml2$curr==0,],modelmat[aml2$curr==1,])
  muc=sum(yc*wi)
  muh=sum(yh*vi)
  muf=(muh+muc+1)/(nc+nh+2)
  va0=(0:50)/50
  ll=lbeta(va0*muh+muc+1,va0*(nh-muh)+nc-muc+1)-lbeta(va0*muh+1,va0*(nh-muh)+1)
  a00=max(va0[ll==max(ll)])
  mu0=(a00*muh+muc+1)/(nc+a00*nh+2)
  fitps=glm(curr~modelmat2,family = 'binomial', weights=c(wi,vi))
  ps=predict(fitps,type="response")[curr==1]
  odd=(1-ps)/ps*vi
  odd=odd/mean(odd)
  muc=sum(yc*wi)
  muh=sum(yh*odd)
  ll=lbeta(va0*muh+muc+1,va0*(nh-muh)+nc-muc+1)-lbeta(va0*muh+1,va0*(nh-muh)+1)
  aps=max(va0[ll==max(ll)])
  mups=(aps*muh+muc+1)/(nc+aps*nh+2)
  Out=rbind(Out,c(mu0,mups,muf,muc/nc))
}

apply(Out,2,mean)
apply(Out,2,var)

plot(density(Out[,1],from=0.8,to=1),ylim=c(0,25),main="",xlab="CR rate at course 2",lwd=1.5)
lines(density(Out[,2],from=0.8,to=1),lty=2,lwd=1.5)
lines(density(Out[,3],from=0.8,to=1),lty=3,lwd=1.5)
lines(density(Out[,4],from=0.8,to=1),lty=4,lwd=1.5)
legend(x=0.8,y=25,legend=c("Dynamic borrowing",
       "Dynamic + IPW borrowing","Full borrowing","No borrowing"),
    lty=1:4)
```

Table 1: Mean bias, variance, MSE and variance ratio(vs no borrowing) of four estimators with no and full borrowing, and dynamic borrowing with and without IPW adjustment for continuous outcome.

| p | b | Method | Bias | Variance | MSE | Var ratio |
|---|---|---|---|---|---|---|
| 5 | 0 | No borrowing | 0.003 | 0.028 | 0.029 | 1.000 |
|   |   | Full borrowing | 0.001 | 0.014 | 0.014 | 0.497 |
|   |   | Dynamic + IPW | 0.003 | 0.026 | 0.026 | 0.901 |
|   |   | Dynamic | 0.003 | 0.022 | 0.022 | 0.770 |
|   | 0.15 | No borrowing | 0.006 | 0.029 | 0.029 | 1.000 |
|   |   | Full borrowing | -0.106 | 0.014 | 0.025 | 0.492 |
|   |   | Dynamic + IPW | 0.029 | 0.024 | 0.025 | 0.843 |
|   |   | Dynamic | -0.022 | 0.026 | 0.027 | 0.898 |
|   | 0.3 | No borrowing | -0.007 | 0.030 | 0.030 | 1.000 |
|   |   | Full borrowing | -0.232 | 0.015 | 0.069 | 0.502 |
|   |   | Dynamic + IPW | 0.013 | 0.024 | 0.024 | 0.814 |
|   |   | Dynamic | -0.041 | 0.032 | 0.034 | 1.070 |
|   | 0.6 | No borrowing | -0.001 | 0.029 | 0.029 | 1.000 |
|   |   | Full borrowing | -0.451 | 0.015 | 0.218 | 0.527 |
|   |   | Dynamic + IPW | 0.001 | 0.027 | 0.027 | 0.946 |
|   |   | Dynamic | -0.018 | 0.030 | 0.031 | 1.048 |
| 10 | 0 | No borrowing | -0.004 | 0.037 | 0.037 | 1.000 |
|   |   | Full borrowing | -0.003 | 0.019 | 0.019 | 0.512 |
|   |   | Dynamic + IPW | -0.004 | 0.035 | 0.035 | 0.949 |
|   |   | Dynamic | -0.004 | 0.028 | 0.029 | 0.770 |
|   | 0.15 | No borrowing | 0.001 | 0.038 | 0.038 | 1.000 |
|   |   | Full borrowing | -0.223 | 0.020 | 0.069 | 0.520 |
|   |   | Dynamic + IPW | 0.020 | 0.033 | 0.033 | 0.861 |
|   |   | Dynamic | -0.039 | 0.040 | 0.041 | 1.042 |
|   | 0.3 | No borrowing | -0.003 | 0.038 | 0.038 | 1.000 |
|   |   | Full borrowing | -0.449 | 0.019 | 0.221 | 0.494 |
|   |   | Dynamic + IPW | 0.001 | 0.035 | 0.035 | 0.930 |
|   |   | Dynamic | -0.026 | 0.040 | 0.041 | 1.067 |
|   | 0.6 | No borrowing | 0.002 | 0.037 | 0.037 | 1.000 |
|   |   | Full borrowing | -0.899 | 0.019 | 0.826 | 0.506 |
|   |   | Dynamic + IPW | 0.001 | 0.037 | 0.037 | 1.001 |
|   |   | Dynamic | -0.009 | 0.037 | 0.037 | 1.013 |

Table 2: Mean bias, variance, MSE and variance ratio(vs no borrowing) of four estimators with no and full borrowing, and dynamic borrowing with and without IPW adjustment for binary outcome.

| $p$ | $b$ | Method | Bias | Variance | MSE | Var ratio |
|---|---|---|---|---|---|---|
| 5 | 0 | No borrowing | 0.000 | 0.005 | 0.005 | 1.000 |
|   |   | Full borrowing | -0.001 | 0.002 | 0.002 | 0.494 |
|   |   | Dynamic + IPW | 0.000 | 0.004 | 0.004 | 0.778 |
|   |   | Dynamic | 0.000 | 0.004 | 0.004 | 0.765 |
|   | 0.3 | No borrowing | -0.001 | 0.005 | 0.005 | 1.000 |
|   |   | Full borrowing | -0.051 | 0.002 | 0.005 | 0.482 |
|   |   | Dynamic + IPW | -0.001 | 0.004 | 0.004 | 0.830 |
|   |   | Dynamic | -0.012 | 0.004 | 0.005 | 0.911 |
|   | 0.6 | No borrowing | -0.001 | 0.005 | 0.005 | 1.000 |
|   |   | Full borrowing | -0.096 | 0.002 | 0.012 | 0.454 |
|   |   | Dynamic + IPW | -0.001 | 0.005 | 0.005 | 0.922 |
|   |   | Dynamic | -0.012 | 0.005 | 0.005 | 1.059 |
|   | 1 | No borrowing | 0.000 | 0.005 | 0.005 | 1.000 |
|   |   | Full borrowing | -0.147 | 0.002 | 0.024 | 0.415 |
|   |   | Dynamic + IPW | -0.001 | 0.005 | 0.005 | 0.994 |
|   |   | Dynamic | -0.006 | 0.005 | 0.005 | 1.047 |
| 10 | 0 | No borrowing | 0.002 | 0.005 | 0.005 | 1.000 |
|   |   | Full borrowing | 0.000 | 0.002 | 0.002 | 0.486 |
|   |   | Dynamic + IPW | 0.001 | 0.004 | 0.004 | 0.801 |
|   |   | Dynamic | 0.001 | 0.004 | 0.004 | 0.765 |
|   | 0.3 | No borrowing | -0.001 | 0.005 | 0.005 | 1.000 |
|   |   | Full borrowing | -0.091 | 0.002 | 0.010 | 0.458 |
|   |   | Dynamic + IPW | -0.001 | 0.004 | 0.004 | 0.888 |
|   |   | Dynamic | -0.013 | 0.005 | 0.005 | 1.051 |
|   | 0.6 | No borrowing | -0.001 | 0.005 | 0.005 | 1.000 |
|   |   | Full borrowing | -0.160 | 0.002 | 0.028 | 0.377 |
|   |   | Dynamic + IPW | -0.002 | 0.005 | 0.005 | 0.988 |
|   |   | Dynamic | -0.006 | 0.005 | 0.005 | 1.041 |
|   | 1 | No borrowing | 0.000 | 0.005 | 0.005 | 1.000 |
|   |   | Full borrowing | -0.215 | 0.001 | 0.048 | 0.305 |
|   |   | Dynamic + IPW | -0.002 | 0.005 | 0.005 | 1.012 |
|   |   | Dynamic | -0.003 | 0.005 | 0.005 | 1.028 |

Table 3: Summary of fitted logistic PS model for probability of being in study AML0531, with raw and IPW weighted covariate differences (AML0531 - AML03P1).

|  | Estimate | Std. Error | z value | Pr(>\|z\|) | Raw Diff. | Weighted Diff. |
|---|---|---|---|---|---|---|
| CNS disease | 0.189 | 0.693 | 0.273 | 0.785 | -0.042 | -0.011 |
| Race | -0.783 | 0.453 | -1.731 | 0.084 | 0.043 | 0.036 |
| log_MRD | 0.059 | 0.132 | 0.449 | 0.653 | -0.044 | -0.043 |
| log_age | -0.082 | 0.153 | -0.538 | 0.591 | 0.057 | -0.209 |
| log_WBC | -0.282 | 0.119 | -2.371 | 0.018 | 0.553 | 0.118 |
| log_BM | -2.703 | 1.942 | -1.392 | 0.164 | 0.242 | 0.068 |
| low_risk | -0.187 | 0.349 | -0.535 | 0.592 | -0.035 | -0.080 |
| high_risk | 1.199 | 0.693 | 1.728 | 0.084 | 0.048 | 0.028 |

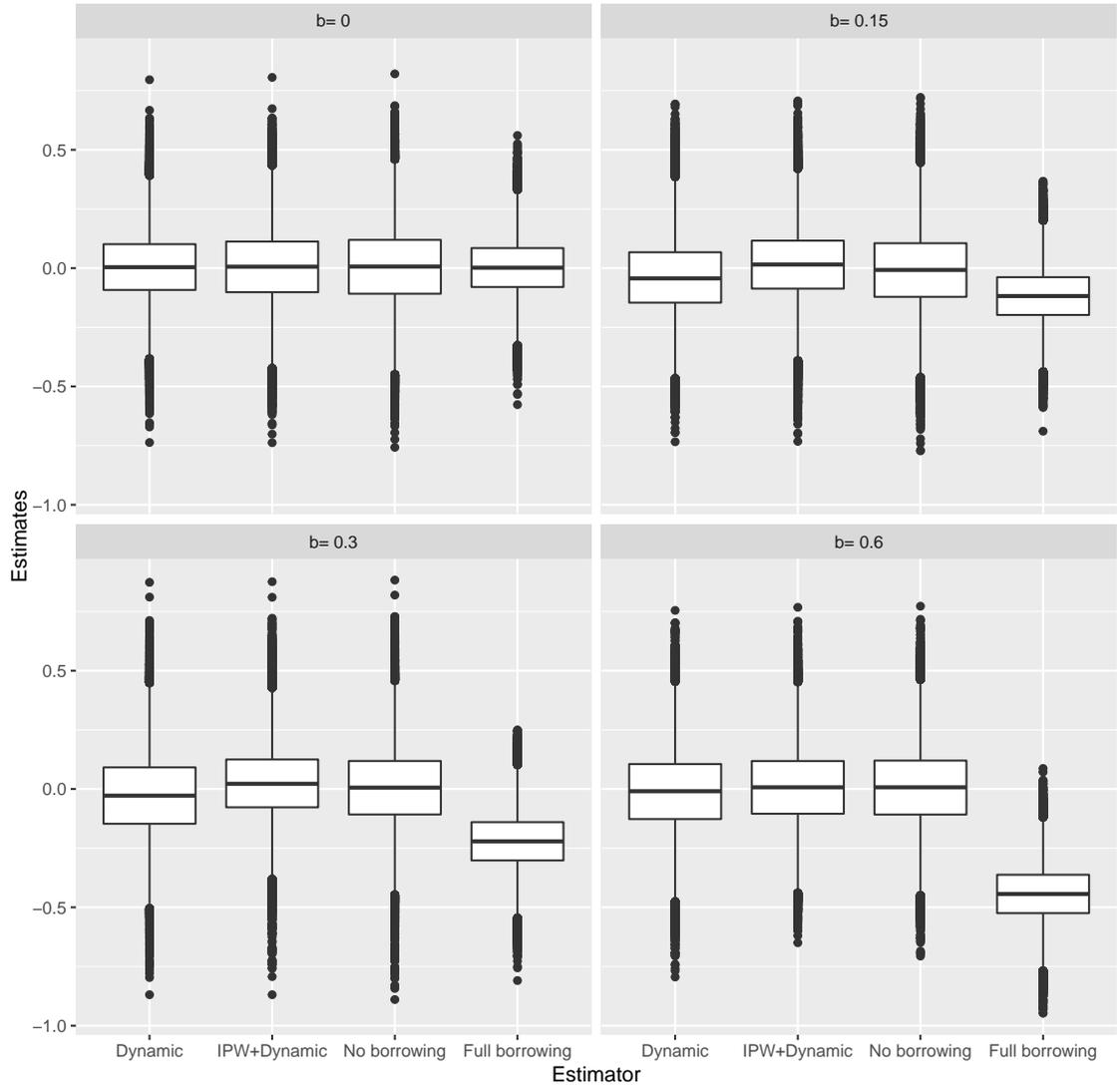

Figure 1: Posterior distributions of $\mu_{cb}$ with and without IPW adjustment by mean difference of covariates between internal and historical controls $E(\mathbf{X}_i|H_i = 0) = 0$ and $E(\mathbf{X}_i|H_i = 1) = -b\mathbf{1}$, compared with those of no borrowing and full borrowing, with normally distributed outcomes. The correct mean outcome under the control treatment is $E(\mathbf{Y}_i(0)) = 0$

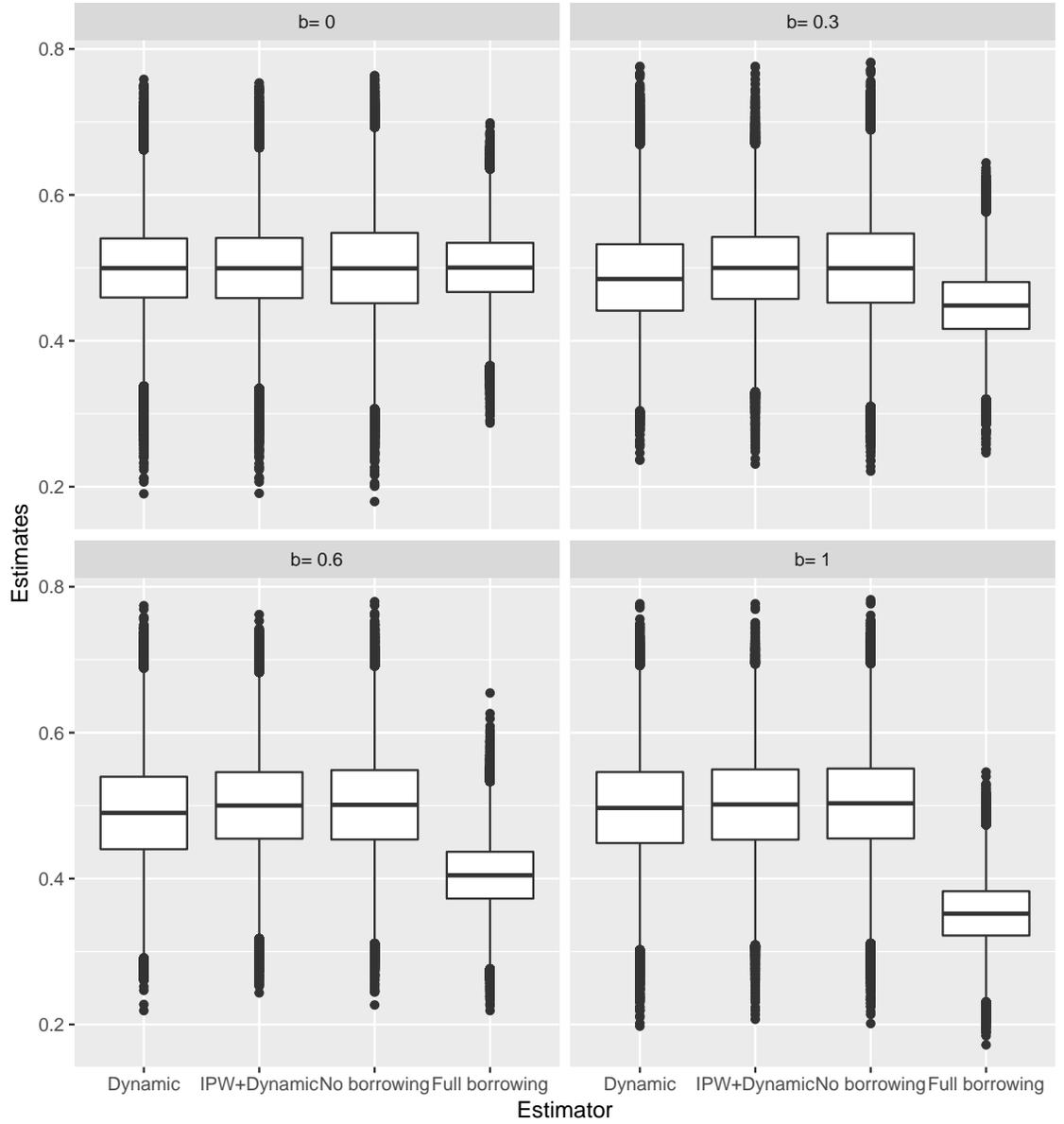

Figure 2: Distribution of $\hat{\mu}_{cb}$ with and without IPW adjustment by mean difference of covariates between internal and historical controls $E(\mathbf{X}_i|H_i = 0) = 0$ and $E(\mathbf{X}_i|H_i = 1) = -b$, compared with those of no borrowing and full borrowing, with binary outcomes. The correct mean outcome under the control treatment is $\mu = 0.5$.

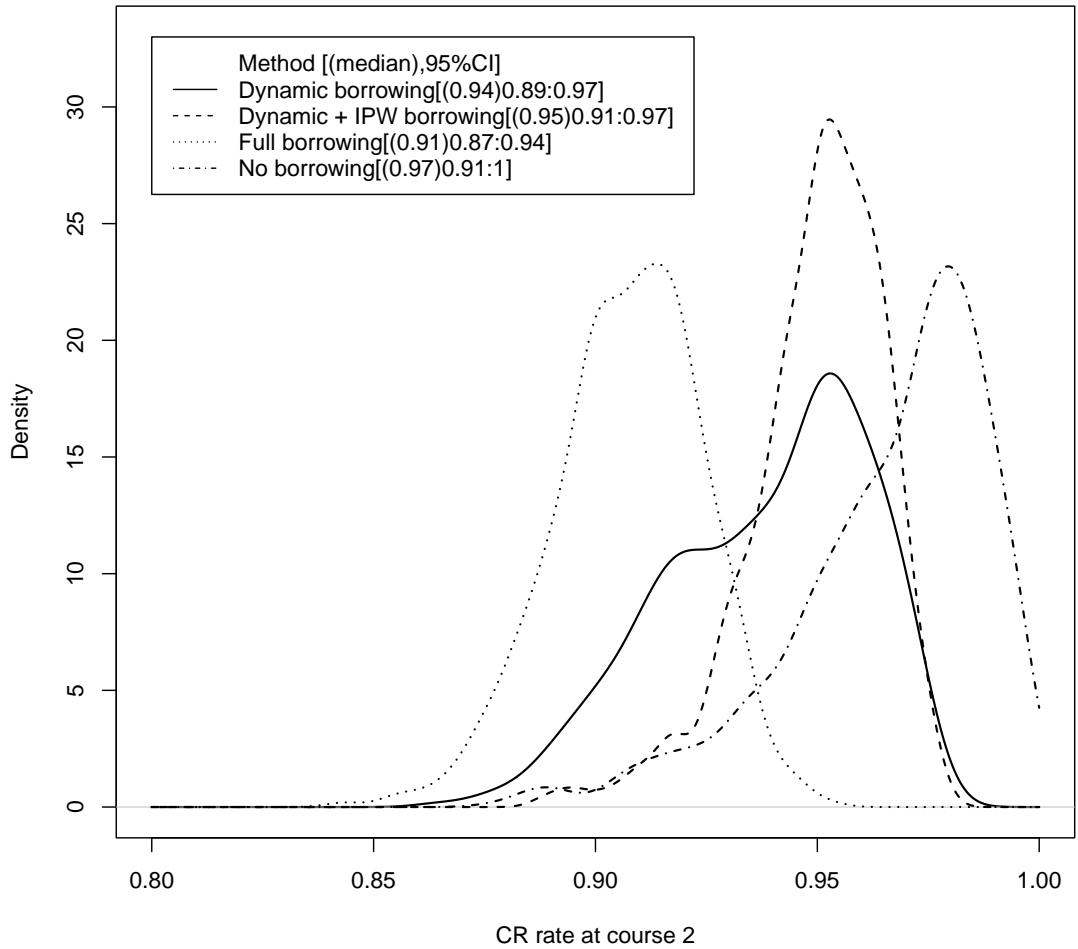

Figure 3: The posterior distribution of CR at the end of course 2 by Bayesian dynamic borrowing with and without IPW adjustment, compared with those with full and no borrowing.